\documentclass[
reprint,
superscriptaddress,
 amsmath,amssymb,
 aps,
prl,
]{revtex4-1}

\usepackage{graphicx}% Include figure files
\usepackage{dcolumn}% Align table columns on decimal point
\usepackage{bm}% bold math
\usepackage[mathlines]{lineno}% Enable numbering of text and display math
\usepackage{mathtools}
\usepackage{amsmath}

\DeclarePairedDelimiter \abs{\lvert}{\rvert}

\begin{document}

\preprint{APS/123-QED}

\title{Coherent Terahertz Spin-Wave Emission Associated with \\ Ferrimagnetic Domain Wall Dynamics}

\author{Se-Hyeok Oh}
\thanks{These two authors contributed equally to this work.}
\affiliation{Department of Nano-Semiconductor and Engineering, Korea University, Seoul 02841, Korea}

\author{Se Kwon Kim}
\thanks{These two authors contributed equally to this work.}
\affiliation{Department of Physics and Astronomy, University of California, Los Angeles, California 90095, USA}

\author{Dong-Kyu Lee}
\affiliation{Department of Materials Science and Engineering, Korea University, Seoul 02841, Korea}

\author{Gyungchoon Go}
\affiliation{Department of Materials Science and Engineering, Korea University, Seoul 02841, Korea}

\author{Kab-Jin Kim}
\affiliation{Department of Physics, Korea Advanced Institute of Science and Technology, Daejeon 34141, Korea}
\affiliation{Institute for Chemical Research, Kyoto University, Kyoto 611-0011, Japan}

\author{Teruo Ono}
\affiliation{Institute for Chemical Research, Kyoto University, Kyoto 611-0011, Japan}

\author{Yaroslav Tserkovnyak}
\affiliation{Department of Physics and Astronomy, University of California, Los Angeles, California 90095, USA}

\author{Kyung-Jin Lee}
\email{kj\_lee@korea.ac.kr}
\affiliation{Department of Nano-Semiconductor and Engineering, Korea University, Seoul 02841, Korea}
\affiliation{Department of Materials Science and Engineering, Korea University, Seoul 02841, Korea}
\affiliation{KU-KIST Graduate School of Converging Science and Technology, Korea University, Seoul 02841, Korea}

\date{\today}% It is always \today, today, but any date may be explicitly specified

\begin{abstract}
We theoretically study the dynamics of ferrimagnetic domain walls in the presence of Dzyaloshinskii-Moriya interaction. We find that an application of a DC magnetic field can induce terahertz spin-wave emission by driving ferrimagnetic domain walls, which is not possible for ferromagnetic or antiferromagnetic domain walls. Dzyaloshinskii-Moriya interaction is shown to facilitate the teraherz spin-wave emission in wide ranges of net angular momentum by increasing the Walker-breakdown field. Moreover, we show that spin-orbit torque combined with Dzyaloshinskii-Moriya interaction also drives a fast ferrimagnetic domain wall motion with emitting terahertz spin-waves in wide ranges of net angular momentum.
%\begin{description}
%\item[PACS numbers] May be entered using the \verb+\pacs{#1}+ command.
% Secondary publications and information retrieval purposes.
% You may use the \texttt{description} environment to structure your abstract;
% use the optional argument of the \verb+\item+ command to give the category of each item.
%\end{description}
\end{abstract}


\maketitle

% \section{\label{sec:level1}First-level heading}
% \subsection{\label{sec:level2}Second-level heading: Formatting}
% \subsubsection{Wide text (A level-3 head)}

%%%%%%%%%%%%%%%%%%%%%%%%%%%%%%%%%%%%%%%%%%%%%%%%%%%%%%%%%%%%%%%%%%%%%%%

In modern communications, information is carried by electromagnetic waves of which frequency is limited to $\approx 0.1$ terahertz (THz), the frequency of oscillating circuits based on high-speed transistors~\cite{Sirtori2002}. On the other hand, semiconductor lasers generate coherent light with the frequency $> 30$ THz~\cite{Beck2002}. Terahertz gap refers to the fact that no relevant technology exists in the frequency range between these two limits ($0.1 \sim 30$ THz). Therefore, it is of critical importance to find relevant physical phenomena that fill in the terahertz gap.

In this respect, antiferromagnets of which resonance frequencies are in the THz ranges~\cite{Nagamiya,Keffer} are of interest~\cite{Jungwirth,Baltz}. It has been reported that coherent THz magnons or spin-waves are generated in antiferromagnets, driven by a laser~\cite{Satoh,Kampfrath} or an electrical current~\cite{Cheng2016,Shiino2016}. However, THz spin-wave excitations by a DC magnetic field are in principle not possible for antiferromagnets as their magnetic moments are compensated on an atomic scale. In this work, we theoretically show that generation of coherent THz spin waves can be achieved by a field-driven domain wall (DW) motion in ferrimagnet/heavy metal bilayers in which the interfacial Dzyaloshinskii-Moriya interaction (DMI) is present. 

As far as the terahertz gap is concerned, this DC field-driven scheme could be beneficial as it allows a low-power operation by avoiding laser-induced or current-induced heating. It is also fundamentally interesting as THz spin-wave emission is caused by an approximately relativistic dynamics of a ferrimagnetic DW. Relativistic kinematics refers to kinematics compatible with the theory of relativity~\cite{Einstein1905}, of which key ingredient is the Lorentz invariance with limiting velocity $c$, the speed of light. When the dispersion of a wave satisfies the Lorentz invariance, a quasiparticle corresponding to the wave follows an analogous relativistic kinematics with replacing the speed of light by the maximum group velocity of the wave. When the velocity of quasiparticle approaches the maximum group velocity, it undergoes the Lorentz contraction and its speed saturates to the limiting velocity. An example of such quasiparticles is an antiferromagnetic DW~\cite{Shiino2016,Haldane1983,Kim2014PRB}. When the DW velocity approaches the maximum spin-wave group velocity, the DW width shirinks with emitting spin-waves~\cite{Shiino2016}. Similarly, the dynamics of a ferrimagnetic DW is also expected to exhibit the features of relativistic kinematics provided that the net magnetization and DMI, which break the Lorentz invariance of the system, are sufficiently ineffective.

%The relativistic DW dynamics is predicted to allow high-frequency spin-wave emission for field-driven ferromagnetic DW motion~\cite{Wang2014} and current-driven antiferromagnetic DW motion~\cite{Shiino2016}. For the former, however, the field-driven relativistic motion is achieved only by assuming very large DW hard-axis anisotropy, comparable to exchange energy, which is not possible in reality. On the other hand, the latter is achieved only by current, thereby causing Joule heating, as field-driven dynamics is not possible for antiferromagnets. Because of zero net magnetic moment, moreover, the creation and detection of antiferromagnetic DWs are technically challenging yet. Therefore, a field-driven THz spin-wave generation is impossible for either ferromagnetic or antiferromagnetic DWs.

In this Letter, we show that such an approximately relativistic DW dynamics is achievable for a class of ferrimagnets, rare earth (RE) and transition metal (TM) compounds, in which the spin moments are antiferromagnetically coupled. As RE and TM elements have different Land{\'e}-g factors~\cite{Jensen1991}, RE-TM ferrimagnets have two distinct temperatures; the magnetic moment compensation point $T_\text{M}$ where net magnetic moment vanishes, and the angular momentum compensation point $T_\text{A}$ where net angular momentum vanishes. For RE-TM ferrimagnets, %spin-transfer torque~\cite{Kaiser2005}, 
resonance~\cite{Binder2006,Stanciu2006}, switching~\cite{Roschewsky2016,Jiang2006,Finley2016,Ueda2016,Mishra2017}, domain wall motion~\cite{Tono2015,Kim2017FiM,Awano2015}, and skyrmion (or bubble domain) motion~\cite{Tanaka2015,Tanaka2017,Kim2017Sky} near these compensation points have been explored experimentally and theoretically. In particular, an experimental observation of a fast field-driven DW motion at $T_\text{A}$ in GdFeCo single-layered ferrimagnets was recently reported~\cite{Kim2017FiM}. This observation reveals two distinguishing features of RE-TM ferrimagnets at $T_\text{A}$. One is that the spin dynamics is antiferromagnetic and thus fast because of zero net angular momentum at $T_\text{A}$. The other is that this fast antiferromagnetic dynamics is achieved by a field because the net magnetic moment is nonzero at $T_\text{A}$ and thus couples with a magnetic field.

%In this Letter, we theoretically study a field-driven motion of a DW in RE-TM type ferrimagnet/heavy metal bilayers and the associated THz spin-wave emission. To that end, 
We begin with deriving the equations of motion for a ferrimagnetic DW based on the collective coordinate approach~\cite{Tretiakov2008}. The dynamics of a general collinear ferrimagnet at sufficiently low temperatures can be described by the following Lagrangian density~\cite{AndreevSPU1980, Kim2017Sky}
\begin{equation}
\label{eq:L}
\mathcal{L} = \rho \dot{\mathbf{n}}^2 / 2 - \delta_s \mathbf{a}[\mathbf{n}] \cdot \dot{\mathbf{n}} - \mathcal{U} \, ,
\end{equation}
where $\mathbf{n}$ is the unit vector along the collinear order, $\rho$ parametrizes the inertia of the dynamics, $\delta_s$ is the spin density in the direction $\mathbf{n}$, $\mathbf{a}[\mathbf{n}]$ is the vector potential generated by a magnetic monopole of unit charge satisfying $\boldsymbol{\nabla}_\mathbf{n} \times \mathbf{a} = \mathbf{n}$, and $\mathcal{U}$ is the potential-energy density. Here, the first term is the spin Berry phase associated with the staggered spin density, which thus appears in the Lagrangian for collinear antiferromagnets; the second term is the Berry phase associated with the net spin density $\delta_s$, which is used to describe the dynamics of uncompensated spins in ferrimagnets. We consider the following potential-energy density:
\begin{equation}
\begin{split}
\label{eq:U}
\mathcal{U} = & A ( \boldsymbol{\nabla} \mathbf{n} )^{2} / 2 - K (\mathbf{n} \cdot \hat{\mathbf{z}} )^{2} / 2 + \kappa (\mathbf{n} \cdot \hat{\mathbf{x}} )^{2} / 2 \\
& + D \hat{\mathbf{y}} \cdot ( \mathbf{n} \times  \partial_{x} \mathbf{n} ) /2 - \mathbf{h} \cdot \mathbf{n} .
\end{split}
\end{equation}
Here, the first term is the exchange energy with $A > 0$; the second term is the easy-axis anisotropy along the $z$ axis with $K > 0$; the third term is the weaker DW hard-axis anisotropy along the $x$ axis with $\kappa > 0$; the fourth term is the interfacial DMI; the last term is the Zeeman coupling with $\mathbf{h} = M \mathbf{H}$, where $M$ is the net magnetization in the direction $\mathbf{n}$. The dissipation can be accounted for by introducing (the spatial density of) the Rayleigh dissipation function, $\mathcal{R} = s_\alpha \dot{\mathbf{n}}^2 / 2$. Here, $s_\alpha$ is a phenomenological parameter quantifying the energy and spin loss due to the magnetic dynamics. For example, in the ferromagnetic regime, i.e., away from $T_\text{A}$, it can be considered as the product of the effective Gilbert damping constant and the net spin density.

The low-energy dynamics of a DW can be described by the two collective coordinates, the position $X(t)$ and the azimuthal angle $\phi(t)$. We consider the Walker ansatz~\cite{Landau1960} for the DW profile: $\mathbf{n}(x,t)=(\sin \theta \cos \phi, \sin \theta \sin \phi, \cos \theta)$, where $\theta = 2 \tan^{-1} \{ \exp [(x-X)/\lambda]\}$ and $\lambda = \sqrt{A / K}$ is the DW width. The equations of motion can be derived from Eqs.~(\ref{eq:L}) and (\ref{eq:U}) in conjunction with the Rayleigh dissipation function:
\begin{eqnarray}
&M& \ddot{X} + G \dot{\phi} + M \dot{X} / \tau = F, \label{Eq:EOM1} \\
&I& \ddot{\phi} - G \dot{X} + I \dot{\phi} / \tau = - \tilde{\kappa} \sin \phi \cos \phi + \tilde{D} \sin \phi, \label{Eq:EOM2}
\end{eqnarray}
where $M = 2 \rho \mathcal{A} / \lambda$ is the mass, $I = 2 \rho \mathcal{A} \lambda$ is the moment of inertia, $G = 2 \delta_\text{s} \mathcal{A}$ is the gyrotropic coefficient, $\tau = \rho /  s_\alpha$ is the relaxation time, $F = 2 h \mathcal{A}$ is the force exerted by an external field, $\tilde{\kappa} = 2 \kappa \lambda \mathcal{A}$, $\tilde{D} = \pi D \mathcal{A} / 2$, and $\mathcal{A}$ is the cross-sectional area of the DW. From Eq.~(\ref{Eq:EOM1}), we obtain the steady-state solution of the DW velocity:
\begin{equation}\label{Eq:Vsteady}
V_\text{DW} = \frac{M \lambda}{ s_\alpha} H,
\end{equation}
where $H$ is the external field applied along the $z$-axis. In this steady state, the DW moves at a constant velocity $V_\text{DW}$ with a constant angle $\phi$. When the field becomes sufficiently strong such that $V_\text{DW}$ exceeds a certain threshold $V_\text{max}$, the DW begins to precess, engendering the phenomenon known as the Walker Breakdown~\cite{Schryer1974,Mougin2007}. The Walker Breakdown field $H_\text{WB}$ can be obtained from Eq.~(\ref{Eq:EOM2}) by
\begin{equation}\label{Eq:HWB1}
H_\text{WB} = V_\text{max} \frac{ s_\alpha}{M \lambda}.
\end{equation}
In the absence of DMI ($D = 0$), the threshold velocity is given by $V_\text{max} = \tilde{\kappa} / 2 G$ and thus $H_\text{WB} = \tilde{\kappa} s_\alpha /  2G M \lambda$. When DMI is much stronger than the DW anisotropy in the $y$ direction, i.e., $|{\tilde D}| \gg {\tilde \kappa}$, then $\abs{V_\text{max}} = \abs{\tilde{D}} / G$. In this strong DMI limit, the Walker Breakdown field is given by
\begin{equation}\label{Eq:HWB2}
H_\text{WB} = \frac{\abs{\tilde D}}{G} \frac{s_\alpha}{M \lambda}= \frac{\pi \abs{D}}{4 \delta_\text{s}} \frac{s_\alpha}{M \lambda}.
\end{equation}
We note that $H_\text{WB}$ is inversely proportional to $G$ and thus to the net spin density $\delta_\text{s}$. As a result, the Walker breakdown is absent at $T_\text{A}$ where the net spin density vanishes, $\delta_\text{s} = 0$. This suppression of the Walker breakdown at $T_\text{A}$ can be understood as a result of decoupling of the DW position $X$ and the angle $\phi$ at $T_\text{A}$~\cite{Kim2017FiM}.

It is worthwhile comparing Eq.~(\ref{Eq:HWB2}) to the Walker breakdown field for ferromagnetic DWs in the strong DMI limit~\cite{Thiaville2012}: $H_\text{WB,FM} = \alpha \pi D_\text{FM} / 2 M_\text{FM} \lambda_\text{FM}$, which can be obtained from Eq.~(\ref{Eq:HWB2}) by taking the ferromagnetic limit. %$s_\alpha = \alpha \delta_s$. 
From this comparison, one finds that in the vicinity of $T_\text{A}$, $H_\text{WB}$ for ferrimagnetic DWs is much larger than that for ferromagnetic DWs because $\delta_\text{s} \approx 0$ and $|M| \ll M_\text{FM}$. Moreover, this very large $H_\text{WB}$ for ferrimagnetic DWs suggests that $V_\text{DW}$ can reach the maximum group velocity of spin-wave more easily without experiencing the Walker breakdown and thus ferrimagnetic DWs can generate THz spin-waves in wide ranges of net angular momentum $\delta_\text{s}$. Finally, the time averaged velocity $\bar{V}$ for a one period far above the Walker Breakdown is given as
\begin{equation}\label{Eq:FarWB}
\bar{V} = \frac{M \lambda}{ s_\alpha + \delta_\text{s}^2 /  s_\alpha} H.
\end{equation}

\begin{figure}[tb]
\begin{center}
\includegraphics[scale=0.37]{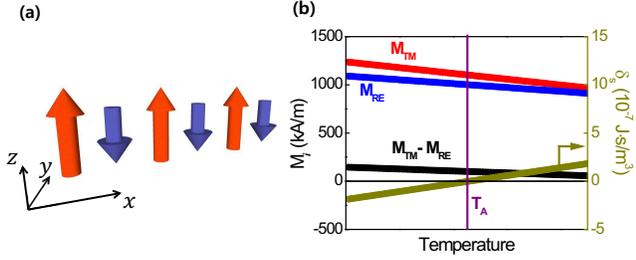}
\caption{
(color online) (a) A schematic illustration of a ferrimagnet in which neighboring spins are coupled antiferromagnetically. (b) The assumed magnetic moments of TM (red) and RE (blue) elements as a function of the temperature $T$. Black symbols represent net magnetic moment (= $M_\text{TM} - M_\text{RE}$), and dark yellow symbols represent net angular momentum $\delta_\text{s}$. Zero $\delta_\text{s}$ corresponds to the angular momentum compensation temperature $T_\text{A}$ (purple). These parameters are used for simulations shown in Figs. 2 and 3.
}
\label{fig:FIG1}
\end{center}
\end{figure}

To verify these theoretical predictions on the DW velocity and THz spin-wave emission, we perform atomistic model calculations~\cite{Shiino2016,Evans2014} for two-sublattice ferrimagnets, which correspond to RE-TM compounds. Two sublattices possess the magnetization $M_1$ and $M_2$, which are coupled by the antiferromagnetic exchange. The spin densities are given by $s_1 = M_1 / \gamma_1$ and $s_2 = M_2 / \gamma_2$, where $\gamma_i = g_i \mu_B / \hbar$ is the gyromagnetic ratio of the lattice $i$, $\mu_B$ is the Bohr magneton, and $g_i$ is the Land\'e-g factor. The parameters in the above descriptions for general ferrimagnets are given by $\delta_s = s_1 - s_2$, $M = M_1 - M_2$, and $s_\alpha = \alpha_1 s_1 + \alpha_2 s_2$, where $\alpha_i$ is the Gilbert damping constant for the lattice $i$. The one-dimensional discrete Hamiltonian that we use for numerical calculations is given by
\begin{eqnarray}\label{Eq:Hamil}
\mathcal{H} &=& A_\text{sim} \sum_{i} \mathbf{S}_{i} \cdot \mathbf{S}_{i+1} - K_\text{sim} \sum_{i} (\mathbf{S}_{i} \cdot \hat{\mathbf{z}})^2 \nonumber  \\
&+& \kappa_\text{sim} \sum_{i} (\mathbf{S}_{i} \cdot \hat{\mathbf{x}})^2 + D_\text{sim} \sum_{i} \hat{\mathbf{y}} \cdot (\mathbf{S}_{i} \times \mathbf{S}_{i+1}) \nonumber \\
&-& g_{i} \mu_\text{B} \mu_{0} \sum_{i} \mathbf{H} \cdot \mathbf{S}_{i},
\end{eqnarray}
where $\mathbf{S}_{i}$ is the normalized spin moment vector at lattice site $i$ [i.e., an even (odd) $i$ corresponds to a RE (TM) atomic site], $A_\text{sim}$, $K_\text{sim}$, $\kappa_\text{sim}$, and $D_\text{sim}$ denote the exchange, easy-axis anisotropy, DW hard-axis anisotropy, and DMI constants, respectively, and $\mathbf{H}$ is the external field. We numerically solve the atomistic Landau-Lifshitz-Gilbert equation:
\begin{eqnarray}\label{Eq:LLG}
\frac{\partial \mathbf{S}_{i}}{\partial t} = - \gamma_{i} \mathbf{S}_{i} \times \mathbf{H_\text{eff,i}} + \alpha_{i} \mathbf{S}_{i} \times \frac{\partial \mathbf{S}_{i}}{\partial t},
\end{eqnarray}
where  $\mathbf{H_\text{eff,i}} = - \frac{1}{M_{i}} \frac{\partial \mathcal{H}}{\partial \mathbf{S}_{i}}$ is the effective field. We use the following simulation parameters: $A_\text{sim}$ = 30 meV, $K_\text{sim}$ = 0.4 meV, $\kappa_\text{sim}$ = 0.2 $\mu$eV, damping constant $\alpha_\text{TM} = \alpha_\text{RE}$ = 0.002, the lattice constant $d$ = 0.4 nm, and Land{\'e} g-factors $g_\text{TM}$ = 2.2 for TM, and $g_\text{RE}$ = 2 for RE element~\cite{Jensen1991}. Figure 1(b) shows the assumed temperature-dependent change in the magnetic moment $M_{i}$ and corresponding $\delta_\text{s}$. For simplicity, we assume other parameters are invariant with temperature.

\begin{figure}[tb]
\begin{center}
\includegraphics[scale=0.43]{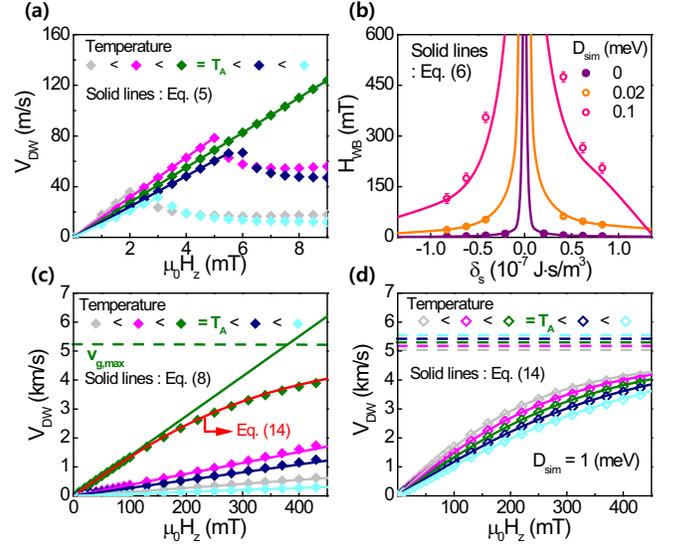}
\caption{
(color online) (a) Domain wall velocity as a function of the external field $H$ in the low-field regime ($\mu_0 H <$ 10 mT). Symbols indicate the calculation results and solid lines indicate Eq.~(\ref{Eq:Vsteady}). (b) Walker Breakdown field $H_\text{WB}$ as a function of net angular momentum $\delta_\text{s}$ at various DMI constants $D_\text{sim}$. Domain wall velocity in the high-field regime for (c) $D_\text{sim} = 0$ and (d) $D_\text{sim}$ = 1 meV. Solid straight lines in (c) indicate Eq.~(\ref{Eq:FarWB}). A red solid line in (c) and solid curved lines in (d) represent relativistically modified solution of domain wall velocity, Eq.~(\ref{Eq:reV}). Horizontal dashed lines in (c) and (d) represent $v_{\rm g,max}$ [Eq.~(\ref{Eq:vgmax})].
}
\label{fig:FIG2}
\end{center}
\end{figure}

Figure 2(a) shows $V_\text{DW}$ for $D = 0$ as a function of $H$. Below $H_\text{WB}$, $V_\text{DW}$ increases linearly with $H$, in agreement with Eq.~(\ref{Eq:Vsteady}) (solid lines). For $H > H_\text{WB}$, the Walker breakdown occurs except for $T = T_\text{A}$ at which $V_\text{DW}$ keeps increasing because of the absence of the Walker breakdown, as explained above. Figure 2(b) shows $H_\text{WB}$ as a function of $\delta_\text{s}$ at various DMIs. Two features are worth mentioning. First, $H_\text{WB}$ diverges at $T_\text{A}$ (i.e., $\delta_\text{s} = 0$). Second, $H_\text{WB}$ for a finite DMI becomes much larger than that for $D = 0$, in agreement with Eq.~(\ref{Eq:HWB1}) (solid lines). Compared to ferromagnets, ferrimagnets exhibit larger $H_\text{WB}$ because the net magnetic moment $M (= M_1 - M_2)$ is small near $T_\text{A}$, as explained above.
%This large increase in $H_\text{WB}$ is caused by a small net magnetic moment ($M_{1} - M_{2}$) near $T_\text{A}$, as explained above. 
Figure 2(c) shows $V_\text{DW}$ in the high field regime for $D = 0$. The numerically obtained values of $V_\text{DW}$ at $T_\text{A}$ (green symbols) deviate from Eq.~(\ref{Eq:FarWB}) (a green solid line), which predicts a linear increase in $V_\text{DW}$ with $H$. This deviation is a manifestation of the approximately relativistic dynamics of ferrimagnetic DW, as we will explain below in detail. For $D = 0$ [Fig. 2(c)], the nonlinear dependence of $V_\text{DW}$ on $H$ appears only at $T_\text{A}$. For $D_\text{sim}$ = 1 meV (corresponding to $D$ = 2 mJ/m$^{2}$), however, the nonlinear dependence appears in all tested ranges of $T$ [or $\delta_\text{s}$; Fig. 2(d)]. It means that the relativistic dynamics occurs in wide ranges of $T$, resulting from the largely enhanced $H_\text{WB}$ near $T_\text{A}$.

For the cases showing the nonlinear dependence of $V_\text{DW}$ on $H$, we observe the spin-wave emission from DW [Fig. 3(a)]. The spin--wave frequency is in THz ranges [Fig. 3(b)]. To elaborate the nonlinear dependence of $V_\text{DW}$ on $H$ and associated THz spin-wave emission, we derive the spin-wave dispersion for ferrimagnets on top of the uniform ground state, $\mathbf{n} = \hat{\mathbf{z}}$, given as
\begin{equation}\label{Eq:SW}
\omega_{\pm} = \frac{\pm \delta_\text{s} + \sqrt{\delta_\text{s}^2 + 4 \rho ( A k^2 + K - h ) }}{2 \rho},
\end{equation}
where the upper (lower) sign corresponds to the right-handed (left-handed) circular mode and $k$ is the wavenumber. %We note that the spin-wave frequency is in THz ranges due to the exchange $A$. 
We note that DMI does not contribute to the spin-wave dispersion as it is effective only for the $y$-component of magnetization, which is negligible for perpendicularly magnetized ferrimagnets.  
From the dispersion [Eq.~(\ref{Eq:SW})], we obtain the maximum spin-wave group velocity $v_{\rm g, max}$ as
\begin{eqnarray}\label{Eq:vgmax}
v_{\rm g, max} = A / ds,
\end{eqnarray}
where $s=(s_1+s_2)/2$. We note that $v_{\rm g, max}$ is indicated by horizontal dashed lines in Fig. 2(c) and (d). With $v_{\rm g, max}$, the nonlinear dependence of $V_\text{DW}$ on $H$ is readily interpreted based on the approximately relativistic kinematics of a DW, similar to the dynamics of an antiferromagnetic DW~\cite{Shiino2016}: $v_{\rm g, max}$ acts as the speed of light~\cite{Kim2014PRB} and the DW width shrinks as $V_\text{DW}$ approaches $v_{\rm g, max}$ via the Lorentz contraction. The Lorentz contraction of DW is described by
\begin{equation}\label{Eq:LC}
\lambda_{\rm DW} = \lambda_{\rm eq} \sqrt{1 - (V_{\rm DW} / v_{\rm g, max})^2 },
\end{equation}
where $\lambda_\text{eq}$ is the equilibrium DW width. The inertial DW mass also varies with the Lorentz factor $1 / \sqrt{1 - (V_\text{DW} / v_{\rm g, max})^{2}}$, i.e., $M = 2 \rho \mathcal{A} / \lambda_{\rm DW}$. Figure 3(c) shows that the DW width decreases as $V_\text{DW}$ approaches $v_{\rm g, max}$ while the inertial mass increases with $V_\text{DW}$. With the Lorentz contraction, we modify Eq.~(\ref{Eq:Vsteady}) relativistically as
\begin{equation}\label{Eq:reV}
V_\text{DW} = v_{\rm g, max} \sqrt{1 - (\lambda_\text{DW} / \lambda_{\rm eq})^2 },
\end{equation}
which is represented by a red solid line in Fig. 2(c) and solid lines in Fig. 2(d). Excellent agreement between numerically obtained $V_\text{DW}$ and Eq.~(\ref{Eq:reV}) confirms the relativistic dynamics of ferrimagnetic DWs.

\begin{figure}[tb]
\begin{center}
\includegraphics[scale=0.41]{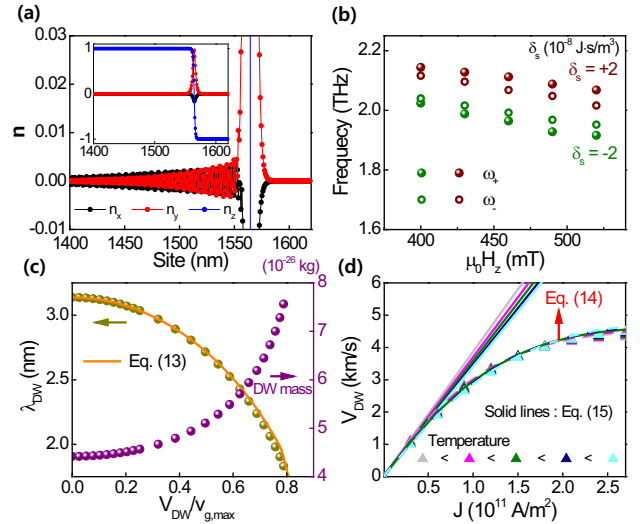}
\caption{
(color online) (a) Configuration of domain wall and spin-waves for the staggered vector $\mathbf{n}$. The inset shows overall shape of domain wall. (b) The frequency of emitted spin-waves as a function of $H$ with finite $\delta_\text{s}$. $\omega_{+}$ and $\omega_{-}$ indicate the right-handed and left-handed spin-wave modes, respectively [see Eq.~(\ref{Eq:SW})]. (c) Domain wall width and mass as a function of $V_\text{DW} / v_{\rm g,max}$ at $T_\text{A}$. The orange solid line indicates Eq.~(\ref{Eq:LC}). (d) Domain wall velocity as a function of the current density $J$ at several temperatures. Symbols represent calculation results and solid lines represent Eq.~(\ref{Eq:vSOT}). Dashed lines represent relativistic modified solutions, Eq.~(\ref{Eq:reV}).
}
\label{fig:FIG3}
\end{center}
\end{figure}

We also show that spin-orbit torque drives a fast relativistic motion of ferrimagnetic DWs. We consider damping-like torque only for simplicity. Using the collective coordinate approach, we obtain the steady-state solution of ferrimagnetic DW velocity as
\begin{equation}\label{Eq:vSOT}
V_{\rm DLT} = \frac{\lambda_\text{DW} \pi s \tilde{B}_{\rm DLT}}{2 \alpha},
\end{equation}
where $\tilde{B}_{\rm DLT} = \hbar \theta_{\rm SH} J / 2 e t_z s_{1} s_{2}$ is the effective field corresponding to the damping-like torque, $\theta_{\rm SH}$ is the effective spin-Hall angle of ferrimagnet/heavy metal bilayer, $J$ is the current density, $e$ is the electron charge, and $t_z$ is the ferrimagnet-thickness. Numerical simulation including the damping-like torque for $D_\text{sim}$ = 1 meV [Fig. 3(d)] shows that spin-orbit torque combined with DMI effect is highly efficient for ferrimagnetic DW motion and the relativistic dynamics occurs for all tested ranges of $T$ (or $\delta_\text{s}$). This spin-orbit-torque-driven ferrimagnetic DW motion also accompanies with THz spin-wave emission (not shown).

Finally, we discuss about the origin of spin-wave emission from ferrimagnetic DWs. Two mechanisms have been proposed: Cherenkov-like process~\cite{Yan} and internal DW structure distortion~\cite{Wang}. The former occurs when DW velocity matches spin-wave phase velocity. In antiferromagnets or ferrimagnets near $T_{\rm A}$, the phase velocity is always higher than the group velocity as one finds from the dispersion [Eq.~(\ref{Eq:SW})]. Therefore, this Cherenkov process is irrelevant to our case. For ferrimagnetic DWs, instead, the spin-waves can be emitted by releasing the DW energy enhanced through the Lorentz contraction as in the case of antiferromagnetic DWs~\cite{Shiino2016}.

In conclusion, we have shown field-driven THz spin-wave emission for ferrimagnetic DWs, which is not possible for ferromagnetic or antiferromagnetic DWs. In ferrimagnet/heavy metal bilayers in which the interfacial DMI arises naturally, the field-driven THz spin-wave emission can be observed in wide ranges of $T$ (or $\delta_\text{s}$), thereby largely enhancing the experimental accessibility to our prediction. Moreover, an in-plane current can also excite THz spin-waves in wide ranges of $T$ (or $\delta_\text{s}$) through the combined effect between spin-orbit torque and DMI. Our finding suggests that ferrimagnetic DWs are potentially useful for high-speed and high-frequency spintronics devices.

\begin{acknowledgments}
K.-J.L. was supported by the National Research Foundation of Korea (NRF) (NRF-2015M3D1A1070465, NRF-2017R1A2B2006119) and by the DGIST R\&D Program of the Ministry of Science, ICT and Future Planning (17-BT-02). S.K.K. and Y.T. were supported by the Army Research Office under Contract No. W911NF-14-1-0016. T.O. was supported by JSPS KAKENHI Grant Numbers 15H05702, 26103002. K.J.K. was supported by NRF-2017R1C1B2009686. G.G. was supported by NRF-2016R1A6A3A1193588.
\end{acknowledgments}

%\bibliography{C:/Users/john/Desktop/Papertex/reff}
\end{document}